\documentclass[12pt]{article}
\usepackage{graphicx}
\newcommand{\beq}{\begin{equation}}
\newcommand{\eeq}{\end{equation}}
\newcommand{\bea}{\begin{eqnarray}}
\newcommand{\eea}{\end{eqnarray}}

\def\sq{{\vbox {\hrule height 0.6pt\hbox{\vrule width 0.6pt\hskip 3pt
   \vbox{\vskip 6pt}\hskip 3pt \vrule width 0.6pt}\hrule height 0.6pt}}}
\begin{document}
\begin{titlepage}
\begin{flushleft}
       \hfill                      {\tt hep-th/0106145}\\
       \hfill                       FIT HE - 01-02 \\
       \hfill                       Kagoshima HE-01-3 \\
\end{flushleft}
\vspace*{3mm}
\begin{center}
{\bf\LARGE Massive vector trapping as a gauge boson on a brane \\ }
%\vspace*{5mm}
\vspace*{12mm}
{\large Kazuo Ghoroku\footnote{\tt gouroku@dontaku.fit.ac.jp} and
Akihiro Nakamura\footnote{\tt nakamura@sci.kagoshima-u.ac.jp}}\\
\vspace*{2mm}

\vspace*{2mm}

\vspace*{4mm}
{\large ${}^1$Fukuoka Institute of Technology, Wajiro, Higashi-ku}\\
{\large Fukuoka 811-0295, Japan\\}
%\vspace*{10mm}
\vspace*{4mm}
{\large ${}^2$Kagoshima University, Korimoto 1-21-35}\\
{\large Kagoshima 890-0065, Japan\\}
\vspace*{10mm}
\end{center}

\begin{abstract}
We propose a mechanism to trap massive vector fields as a photon
on the Randall-Sundrum brane embedded in the five 
dimensional AdS space. This localization-mechanism of the photon
is realized by considering a brane action, to which a quadratic potential 
of the bulk-vector fields is added. We also point out that this potential 
gives several constraints on the fluctuations of the vector fields 
in the bulk space.  

\end{abstract}
\end{titlepage}

\section{Introduction}

People expect that our four dimensional world would be embedded
in the higher dimensional space-time where the geometry would be
determined by the superstring theory. The Randall-Sundrum brane-model
(RS brane) ~\cite{RS1,RS2}
would be a probable candidate of such a simplified theory with more
extra-dimensions.
It is possible to consider such that the RS brane is embedded at some
point of the transverse coordinate in the five-dimensional anti-de Sitter
(AdS$_5$) space. And AdS$_5$ is 
realized near the horizon of the background geometry constructed
by the stack of many D3-branes.

In order to consider this RS brane as our four-dimensional
world, it would be necessary to confine all observed fields 
in this brane. Up to now, it is known that the gravity and
scalars can be trapped on the brane of positive tension \cite{RS2,BG},
and fermions are localizable on the one of negative tension \cite{GrN,CHN,BG}
due to the gravitational force coming from the
background configuration, AdS$_5$. However no one knows how the gauge bosons
can be trapped on the RS brane by the same situation. 

Some ideas \cite{DS,DGS,DFKK,DLS}
for the localization of the gauge fields have been proposed
without relating the mechanism to the gravitational force only.
Another interesting mechanism has been proposed in \cite{Oda}, where
a special mass term has been introduced in the bulk action through
three-form field. As a result, the zero-mode of the photon becomes
localizable on the brane.
However, a resultant four-dimensional action contains a mass term
of the gauge boson characterized by the introduced topological mass. 
In this sense, this mechanism should be modified to a more reasonable one.

The purpose of this paper is to propose a new localization-mechanism of the 
gauge field by considering a massive vector in the bulk and a 
slightly modified 
brane-action. The modification of the brane-action 
is performed by adding a localized potential 
of the vector fields which are living in the bulk.  
The idea to add such a potential of the bulk field 
is also seen in \cite{GW} for the case of scalar fields, 
and its possible origin might be found in 
the quantum corrections in the bulk \cite{GGH}.  
Here the 
potential on the brane is given by a quadratic form of this vector field, 
which can be regarded as a localized mass-term of the vector.  
Then we can show that the zero-mode, which means zero four-dimensional 
mass of this mode, of the vector is localized on the brane as a gauge 
field. In other words, the broken gauge-symmetry in the bulk is restored 
on the brane of reduced dimensions. Then our mechanism would be inverse of 
the one given in \cite{Oda}, where 
the bulk gauge-symmetry seems to be broken on the brane due to the 
induced mass-term of the localized vector-field.  
We should point out another point that the solutions for the fluctuations 
of the massive vector in AdS$_5$ background are solved by including 
odd functions with respect to the coordinate perpendicular to the brane.  

In the next section, our localization-mechanism is shown by solving the field 
equations of the vector fields according to the parallel method used 
in \cite{GN,11}. In the section three, this result is further assured through 
the study of the Green function of the vector in the bulk space.  
Final section is devoted to the summary.

\section{Localized state of the vector }

Here we start from the following effective action, 
\beq
 S= S_{\rm gr}+S_{\rm A}.
\eeq
The first term denotes the gravitational part, 
\beq
    S_{\rm gr} = {1\over 2\kappa^2}\Bigg\{
      \int d^5X\sqrt{-G} (R + \Lambda)
          +2\int d^4x\sqrt{-g}K\Bigg\}-{\tau\over 2}\int d^4x\sqrt{-g}, 
                  \label{action}
\eeq
where $K$ is the extrinsic curvature on the boundary, and $\tau$ represents 
the tension of the brane. The background configuration with RS brane, 
\beq
 ds^2= e^{-2|y|/L}\eta_{\mu\nu}dx^{\mu}dx^{\nu}
           +dy^2  \, \label{metrica}
\eeq
is determined by this action, where $\eta_{\mu\nu}=(-,+,\cdots,+)$.  
Here $\tau=6/(L\kappa^2)$ and 
$L=\sqrt{6/\kappa^2\Lambda}$ which denotes the radius 
of five-dimensional AdS space.  
The coordinates parallel to the brane are denoted by $x^{\mu}$ 
and $y$ is the coordinate transverse to the brane.  

The second part $S_{\rm A}$ denotes the action for the massive vector, 
which is denoted by $A_M(x,y)$, 
\bea
 S_{\rm A} &=& \int d^5X\sqrt{-G}\left(-{1\over 4}G^{MN}G^{PQ}F_{MP}F_{NQ}
               - {1\over 2}M^2 G^{MN}A_M A_N\right)   \cr
           && -c\int d^4x{1\over 2}\sqrt{-g}g^{MN}A_M A_N \, ,\label{SA}
\eea
where the second integral is defined on $X^5=y=0$.  
The parameter $M$ and $c$ denote the bulk mass of the vector and 
the coupling of the vector potential and the brane respectively.  
Here no field other than the vector and gravity is considered, 
so we ignore to consider 
the origin of the mass ($M$) of $A_M$ and the gauge symmetry 
expected before the generation of $M$ in the bulk. And this action 
is the starting point.  

Here we concentrate our attention on the behaviors of the 
vector-fields fluctuation around the background (\ref{metrica}) 
according to the analysis of the massive scalar given in \cite{GN}.  
Then the field equation of $A_M$ is given as 
\beq
 {1\over\sqrt{-G}}\partial_A[\sqrt{-G}
  (G^{AB}G^{CD}-G^{AC}G^{BD})\partial_B A_C] -
           [M^2 + c\delta(y)]G^{DB}A_B =0 . \label{eq1}
\eeq
This equation can be solved by separating to the one of $A_y$ and $A_{\mu}$, 
the transverse and parallel parts to the brane. The parallel part 
$A_{\mu}$ is further separated to the transverse and longitudinal 
parts with respect to the four-dimensional momentum on the brane, 
\beq
  A_{\mu} = A_{\mu}^T + A_{\mu}^L ,
\eeq
where 
$A_{\mu}^T
=(\eta_{\mu\nu}-\partial_{\mu}\partial_{\nu}/\sq)\eta^{\nu\rho}A_{\rho}$
and $A_{\mu}^L=(\partial_{\mu}\partial_{\nu}/\sq)\eta^{\nu\rho}A_{\rho}$
with $\sq=\eta^{\mu\nu}\partial_{\mu}\partial_{\nu}$.  
%%%%%%%%%%%%%  Added 1  %%%%%%%%%%%%%%%%%%
The projection-operators for $A_{\rho}$ 
presented here are useful in the case of non-zero 
eigenvalue for $\sq$.  
For the case of zero-eigenvalue, this projection is not necessary 
since $A_{\mu}^L$ is absent if the gauge fixing term is not considered.  
%%%%%%%%%%%%%  Added 1 %%%%%%%%%%%%%%%%%%

%%%%%%%%%%%%%  Added 2 %%%%%%%%%%%%%%%%%%
We notice that it is necessary to consider even and odd functions both for 
$y\to -y$ in solving Eq.~(\ref{eq1}). This is a different point from the 
case of the massless vector fields.  
Here we restrict to the following two simple cases; $A_{\mu}^T$ is even, i.e., 
$A_{\mu}^T(x,y) = A_{\mu}^T(x,|y|)$, and either $A_{\mu}^L$ or $A_y$ is 
odd. Namely, 
\beq
  A_{\mu}^L(x,y) = {\rm sgn}(y)f_{\mu}(x,|y|)  , \quad 
        A_y(x,y) = f_y(x,|y|),          \label{casea}
\eeq
or 
\beq
      A_y(x,y) = {\rm sgn}(y)f_{y}(x,|y|)  , \quad 
       A_{\mu}^L(x,y) = f_{\mu}(x,|y|).   \label{caseb}
\eeq
Here ${\rm sgn}(y)$ is defined as ${\rm sgn}(y)=|y|/y$ for $y\neq 0$ 
and ${\rm sgn}(y)=0$ for $y=0$.
In both cases, we obtain from Eq.~(\ref{eq1}) by operating $\partial_D$ 
on it the following relation 
\beq 
  \eta^{\mu\nu}\partial_{\mu}f_{\nu}=-e^{-2|y|/L}(f_y'-{4\over L}f_y) ,
                    \label{bian}
\eeq
where $f_y'=df_y/d|y|$. Here we solve (\ref{eq1}) with respect to 
$A_{\mu}^T$ and $A_y$ firstly, then $A_{\mu}^L$ is obtained from 
(\ref{bian}) in terms of the $A_y$ solved.  
%%%%%%%%%%%%%  Added 2 %%%%%%%%%%%%%%%%%%

Using (\ref{eq1}) and (\ref{bian}), we obtain the equations for 
$A_{\mu}^T$ and ${A}_{y}$. In order to write these equations in the form 
of one-dimensional ``Schr\"odinger equation'', 
$A_{\mu}^T$ and ${A}_{y}$ are replaced in the form, 
\beq
  A_{\mu}^T=(|z|/L+1)^{1/2}\hat{\psi}(x,z), \qquad
   A_{y}=(|z|/L+1)^{5/2}\hat{\psi}(x,z),
\eeq
where $z={\rm sgn}(y)L(e^{|y|/L}-1)$. Further imposing 
$\sq\hat{\psi}(x,z)=m^2\hat{\psi}(x,z)$ for $\hat{\psi}(x,z)$, 
where $m$ denotes the four-dimensional mass of the bulk fields.  
Then we obtain 
\beq
 [-\partial_z^2+V(z)]\hat{\psi}(x,z)=m^2\hat{\psi}(x,z) , \ \label{eq3}
\eeq
where 
\beq
 V(z)={a \over (|z|/L+1)^2}-{b}\delta(z), \qquad 
a={3\over 4L^2}+M^2 . \label{poten}
\eeq
The parameter $b$ is given by $b=1/L-c$ for $A_{\mu}^T$. While 
$b=-3/L-c$ and $b=-3/L-4M^2/c$ for ${A}_{y}$ of (\ref{casea}) and 
(\ref{caseb}) respectively.  
In (\ref{eq3}), the $x$-dependent part can be factored out, so we 
concentrate on its $z$-dependence hereafter.  

For $A_{\mu}^T$, the equation (\ref{eq3}) 
can be solved with $b=1/L-c$ and the following boundary condition, 
\beq
  \partial_z\hat{\psi}(z)|_{z=0}= -{1\over 2}(1/L-c)\hat{\psi}(z)|_{z=0} ,
              \label{bound3}
\eeq
which is required from the equation (\ref{eq3}) due to the existence 
of the $\delta$-function in the potential. It is easy to see that this 
condition is consistent with the one obtained from (\ref{eq1}) for $A_{\mu}^T$. 
Then the solution for $m>0$ is obtained as 
\beq
  \hat{\psi}(z)=N (|z|/L+1)^{1/2}[J_{\nu}(m[|z|+L])+\alpha
               Y_{\nu}(m[|z|+L])]      \label{soltr}
\eeq
where $\nu=\sqrt{1+M^2L^2}$, 
$N$ denotes the normalization factor and 
\beq
  \alpha=-{mJ_{\nu}'(mL)+(1/L-c/2)J_{\nu}(mL) \over
              mY_{\nu}'(mL)+(1/L-c/2)Y_{\nu}(mL)}.
\eeq
But we know that this mode does not localize on the brane. It is 
seen from the equation (\ref{eq3}). For $1/L>c$ and $a>0$, 
we obtain the so-called 
volcano-type potential which is necessary for the localization of the 
mode given by the solution of this equation. However any mode with 
$m>0$ would decay into the bulk with a finite life-time as shown in 
\cite{GN,DRT}. The modes with $m>0$ are identified with the continuous 
Kaluza-Klein (KK) modes.  
Then the localizable and stable state is restricted to 
the ``zero-mode'' of $m=0$ \cite{GN}.  

The normalizable solution of zero-mode is given by 
\beq
  \hat{\psi}(z)=N(|z|/L+1)^{1/2-\nu},  \label{renor}
\eeq
In this case, the parameter $c$ is determined from the boundary 
condition (\ref{bound3}) as, 
\beq
   c=-{{2(\nu-1)} \over L}. \label{cval}
\eeq
Then, this mode is localized on the brane as a massless vector boson, 
namely as the gauge boson, when the parameter $c$ is chosen as above.

%%%%%%%%%%%%%%%%%%%%%%%%  Modified  %%%%%%%%%%%%%%%%%%%%
Next, we examine 
$A_y$. Its solution can be obtained similarly to the case of $A_{\mu}^T$ 
by solving (\ref{eq3}) with $b=-3/L-c$ and the boundary condition, 
\beq
  \partial_z\hat{\psi}(z)|_{z=0}= {1\over 2}(3/L+c)\hat{\psi}(z)|_{z=0} ,
              \label{bound4}
\eeq
for the solution of (\ref{casea}).
In the case of (\ref{caseb}), the boundary condition is obtained by 
replacing the second term $c$ in the parenthesis of (\ref{bound4}) 
by $4M^2/c$. In any case, the explicit form of the solution could be given 
in a similar form to (\ref{soltr}) with a different $\alpha$ 
which is determined by the boundary condition.  
We do not give it since it is not important here.  

Interesting point 
to be examined for this solution is that we could expect 
zero-mode bound state for $b>0$ or $c<-3/L$, where attractive 
$\delta$-function potential exists.  
The necessary condition to realize this bound state is obtained 
in a similar way as in the case of $A_{\mu}^T$. The normalizable 
solution of the zero-mode is given by the same form with (\ref{renor}) 
in this case also, and we obtain (\ref{cval}) as the condition for the 
case of (\ref{caseb}).  
This fact seems to imply that we can observe both massless vector 
and scalar, which is identified with $A_y$ here, on the brane in 
this case.  However scalar field can not be seen on the brane because 
of the factor ${\rm sgn}(y)$ which vanishes at $y=0$, just on the brane.  

For the case of (\ref{casea}), the condition is obtained as follows, 
\beq
   c=-{{2(\nu+1)} \over L}.  \label{cval3}
\eeq
In this case, the value of $c$ given by (\ref{cval}) does not satisfy 
the above equations at any $\nu$, 
so the scalar zero-mode can not be bounded when 
vector zero-mode is bounded and vice versa.  

The second point to be studied is how is about the bound state of the
tachyonic scalar, the state with $m^2<0$, when the zero-mode of 
$A_{\mu}^T$ is bounded.  
Since only one bound state is expected for the case of 
$\delta$-function potential, there is no tachyonic bound state 
for the case of $A_{\mu}^T$ and $A_y$ for (\ref{caseb}).  
While for $A_y$ of (\ref{casea}), we can expect the tachyonic 
bound state since the value of $b$ is smaller than that 
of $A_{\mu}^T$ or the one of $A_y$ for the case of (\ref{caseb}).  
The values of $b$ for the latter two cases are the same, and zero-mode 
is bounded for both cases. If there exists a tachyonic bound state
of $A_y$ for (\ref{casea}), 
its position at $m=i|m|$ is given by solving the following 
equation \cite{GN}, 
\beq
   L|m| K_{\nu-1}(L|m|)+2K_{\nu}(L|m|)=0.  \label{cval4}
\eeq
It is easy to assure that there is no solution for 
$|m|>0$, where both terms in the left hand side of the 
equation (\ref{cval4}) are positive for any 
$\nu$.  
Then we can realize the situation where only ``photon'' is bounded 
on the brane without any tachyonic state.  

As for the tachyonic bound state for $A_{\mu}^T$ or the one of $A_y$ 
for the case of (\ref{caseb}), it appears when  $c$ is taken as an 
appropriate negative value. For example, we can find it for $c=-2/L$ 
and $\nu<2$.  We do not consider these cases here.
%%%%%%%%%%%%%%%%%%%  Modified  %%%%%%%%%%%%%%%%%%

The effective action for the localized zero-mode of $A_{\mu}^T$ can 
be written by denoting it as $A_{\mu}(x,y)=a_{\mu}(x)u(y)$ and 
substituting the solution obtained above for $u(y)$, which is given by 
\beq
 u(y)=e^{-(\nu-1)y/L}
\eeq
where we set $u(0)=1$ since the normalization can be absorbed into 
$a_{\mu}(x)$. Then the effective action 
is obtained as 
\beq
    \int d^4x \left(-{1\over 4}f_{\mu\nu}f^{\mu\nu}\right), \label{aeff}
\eeq
where the suffices are contracted by $\eta_{\mu\nu}$ and 
$f_{\mu\nu}=\partial_{\mu}\tilde{a}_{\nu}-\partial_{\nu}\tilde{a}_{\mu}$.  
Here 
\beq
 \tilde{a}_{\mu}(x)=\sqrt{\int_0^{\infty}dy u(y)^2}{a}_{\mu}(x)
                   =\sqrt{{L\over 2(\nu-1)}}{a}_{\mu}(x).
                 \label{norm}
\eeq
The above integral with respect to $y$ is finite for $\nu>1$ or $M^2>0$, 
which is the case considered here.  
%%%%%%%%%%%%%%%%%%%%%%%%%%%%%%%%%%%%%%%%%%   added %%%%%%

In any case, we can see that the zero mode of the bulk vector is 
localized on the brane as a gauge boson when a quadratic form of 
potential of this vector is added on the brane. In the next section, 
we see this point from the Green function for the vector fields.

%%%%%%%%%%%%%%%%%%   Vector propagator   %%%%%%%%%%%%%%%%%%%%%%%%%
\section{Vector propagator}
In this section we analyze the Green function of massive vector in the 
$d+1$ dimensional AdS space to see its effective propagator 
observed in $d$ dimensional flat space of the brane.  
After obtaining the Green function, we go back to the case of $d=4$.  
We work in the brane background of the AdS metric (\ref{metrica}). 

The action integral for free 
massive vector $A_M(X)$ with external source $J_M(X)$ is given by 
\beq
 S_{d+1} = S_{A\,d+1}-\int d^{d+1}X\sqrt{-G}G^{MN}A_M J_N, \label{Sd+1}
\eeq
where $S_{A\,d+1}$ is the $d+1$ dimensional version of $S_A$ 
defined by (\ref{SA}).  
Then the Green function $\Delta_{MN}(X,X')$ is defined by 
\beq
D^{BN}(X)\Delta_{NP}(X,X')
  ={1\over{\sqrt{-G}}}\delta^B{}_P\delta^{d+1}(X-X'), \label{gfdef1}
\eeq
\bea
D^{BN}\Delta_{NP}&=&{1\over{\sqrt{-G}}}\partial_A[\sqrt{-G}
  (G^{AM}G^{BN}-G^{AN}G^{BM})\partial_M\Delta_{NP}]  \cr
&&  -[M^2+c\delta(y)] G^{BN}\Delta_{NP}. \label{gfdef2}
\eea
From this definition it follows the relation 
\beq
{1\over{\sqrt{-G}}}\partial_B({\sqrt{-G}}[M^2+c\delta(y)]G^{BN}\Delta_{NP})
=-{1\over{\sqrt{-G}}}\partial_P\delta^{d+1}(X-X'), \label{Bianp1}
\eeq
as a consequence of the identity 
\beq
\partial_B\partial_A[\sqrt{-G}(G^{AM}G^{BN}-G^{AN}G^{BM})
\partial_M\Delta_{NP}]=0. \label{ident}
\eeq
In (\ref{gfdef1}) and (\ref{Bianp1}), it should be understood as 
$\delta^{d+1}(X-X')=\delta^d(x-x')\delta(y-y')$.

Let us change variables from the coordinate $X^M=(x^\mu, y)$ 
to the coordinate $X^M=(x^\mu,z)$, where $z=L\exp(|y|/L)$, 
which is different from $z$ used in the previous section.  In terms of 
the coordinate $X^M=(x^\mu,z)$, AdS metric is written as 
\beq
  ds^2 = G_{MN}dX^M dX^N
       = {{L^2}\over{z^2}}(\eta_{\mu\nu}dx^\mu dx^\nu + dz^2).
\eeq
The propagators are assumed to be
\bea
\Delta_{\mu\nu}(x,y;x',y') &=& \Delta_{\mu\nu}(x,z;x',z'),\\
\Delta_{y \nu}(x,y;x',y') 
&=& {\rm sgn}(y){z\over L}\Delta_{z \nu}(x,z;x',z'),\\
\Delta_{\mu y}(x,y;x',y') 
&=& \Delta_{\mu z}(x,z;x',z'){{z'}\over L}{\rm sgn}(y'),\\
\Delta_{yy}(x,y;x',y') 
&=& {\rm sgn}(y){z\over L}\Delta_{zz}(x,z;x',z'){{z'}\over L}{\rm sgn}(y'),
\eea
These forms are inspired from the general coordinate transformation
and the possibility considered in the previous section that $A_y(x,y)$
is odd function of $y$.  

The components $D^{BN}$ in the definition (\ref{gfdef1}) 
are given by the followings; 
\bea
D^{\beta\nu}&=& \left({z\over L}\right)^4[\eta^{\beta\nu}D_1
  -\eta^{\alpha\nu}\eta^{\beta\mu}\partial_\alpha\partial_\mu]
  +\delta(z-L)\eta^{\beta\nu}(2\partial_z-c), \\
D^{\beta z} &=& -\left({z\over L}\right)^4\eta^{\beta\mu}\partial_\mu
  \left(\partial_z - {{d-3}\over z}\right)
  -\delta(z-L)2\eta^{\beta\mu}\partial_\mu, \\
D^{z\nu} &=& -\left({z\over L}\right)^4\eta^{\alpha\nu}\partial_\alpha
  \partial_z, \\
D^{zz} &=& \left({z\over L}\right)^4\left[\sq-{{(ML)^2}\over{z^2}}\right], \\
D_1 &\equiv& \partial_z^2-{{d-3}\over z}\partial_z-{{(ML)^2}\over{z^2}}+\sq. 
\eea
Here, in (\ref{gfdef1}), it should be understood as $B,N=\mu,z$ and 
$\delta^{d+1}(X-X')=\delta^d(x-x')\delta(z-z')$.  
As for the relation (\ref{Bianp1}) reads in components as 
\bea
&&M^2\left[\eta^{\beta\nu}\partial_\beta\Delta_{\nu P}
+\left(\partial_z-{{d-1}\over z}\right)\Delta_{zP}\right]
+\delta(z-L)(c\eta^{\beta\nu}\partial_\beta\Delta_{\nu P}+2M^2\Delta_{zP}) \cr
&=&-\left({z\over L}\right)^{d-1}\partial_P\delta^{d+1}(X-X'). \label{Bianp2}
\eea
Here also it should be understood as $P=\mu,z$ and 
$\delta^{d+1}(X-X')=\delta^d(x-x')\delta(z-z')$.  

The defining equations (\ref{gfdef1}) for the Green functions 
are solved off brane ($z\ne L$) as follows.  
(i) With the aid of relations (\ref{Bianp2}), 
$\eta^{\beta\nu}\partial_\beta\Delta_{\nu P}$ are written in terms of 
$\Delta_{zP}$.  
(ii) Then closed equations are obtained for $\Delta_{zz}$ 
and $\Delta_{z\rho}$ by themselves and they are easily solved.  
(iii) As for $\Delta_{\alpha z}$ and $\Delta_{\alpha\rho}$, 
the equations become inhomogeneous equation again with the aid of 
relations (\ref{Bianp2}).  Since the inhomogeneous terms 
for $\Delta_{\alpha z}$ ($\Delta_{\alpha\rho}$) is already solved 
$\Delta_{zz}$ ($\Delta_{z\rho}$) and $\delta(z-z')$, they are solved 
with as yet undetermined homogeneous solutions.  
(iv) The undetermined homogeneous solutions are determined 
in such a way that $\Delta_{MN}$ satisfy the relation 
(\ref{Bianp2}).  The resulting solutions are given by the followings; 
\bea
\Delta_{\alpha\rho} &=& \Delta_{\alpha\rho}^T + \Delta_{\alpha\rho}^L,
 \label{psol1} \\
\Delta_{\alpha\rho}^T &=& \left({{zz'}\over{L^2}}\right)^{d/2-1}
  \int{{d^d p}\over{(2\pi)^d}}e^{ip(x-x')}\left(\eta_{\alpha\rho}
  -{{p_\alpha p_\rho}\over{p^2}}\right)\hat\Delta_1(p,z,z'), 
 \label{psol2}\\
\Delta_{\alpha\rho}^L &=& -{1\over{M^2}}\left({{zz'}\over{L^2}}\right)^{d/2}
  \int{{d^d p}\over{(2\pi)^d}}e^{ip(x-x')}{{p_\alpha p_\rho}\over{p^2}}
  \Biggl[{L\over{z'}}\delta(z-z') \cr
&&+\left(\partial_z-{{d/2-1}\over z}\right)
  \left(\partial'_z-{{d/2-1}\over{z'}}\right)\hat\Delta_2(p,z,z')\Biggr],
      \label{psol3} \\
\Delta_{\alpha z} &=& -{1\over{M^2}}\partial_\alpha
  \left(\partial_z-{{d-1}\over z}\right)\Delta_2(X,X'), \label{psol4}\\
\Delta_{z\rho} &=& -{1\over{M^2}}\left(\partial'_z-{{d-1}\over{z'}}\right)
  \partial'_\rho\Delta_2(X,X'), \label{psol5}\\
\Delta_{zz} &=& -{1\over{M^2}}\left({{zz'}\over{L^2}}\right)^{{{d-1}\over 2}}
  \delta^{d+1}(X-X')+{{\sq}\over{M^2}}\Delta_2(X,X'), \label{psol6}
\eea
where $\partial'_z$ ($\partial'_\alpha$) denotes differentiation with 
respect to $z'$ ($x^{\prime\alpha}$).  
The scalar-type propagators $\Delta_1(X,X')$ and $\Delta_2(X,X')$ 
are defined by the followings;
\bea
&&D_1(X)\Delta_1(X,X')=\left({{z'}\over L}\right)^{d-3}\delta^{d+1}(X-X'), \\
&&D_2(X)\Delta_2(X,X')=\left({{z'}\over L}\right)^{d-1}\delta^{d+1}(X-X'), \\
&&D_2 \equiv \partial_z^2-{{d-1}\over z}\partial_z-{{-(d-1)+(ML)^2}\over{z^2}}
     +\sq, \\
&&\Delta_1(X,X')=\left({{zz'}\over{L^2}}\right)^{d/2-1}
  \int{{d^d p}\over{(2\pi)^d}}e^{ip(x-x')}\hat\Delta_1(p,z,z'), \\
&&\Delta_2(X,X')=\left({{zz'}\over{L^2}}\right)^{d/2}
  \int{{d^d p}\over{(2\pi)^d}}e^{ip(x-x')}\hat\Delta_2(p,z,z'), \\
&&\left[\partial_z^2+{1\over z}\partial_z+\left(q^2-{{\nu^2}\over{z^2}}\right)
  \right]\hat\Delta_i(p,z,z')={L\over{z'}}\delta(z-z'),\qquad (i=1,2) 
 \label{besseq}\\
&&q^2=-p^2,\qquad  \nu\equiv[(d/2-1)^2+(ML)^2]^{1/2}
\eea
It should be noted that the $\delta$-function singularity is canceled 
by the second term in the square bracket of (\ref{psol3}) and that 
$\hat\Delta_1$ only contributes to transverse parts while 
$\hat\Delta_2$ contributes to longitudinal parts.  
It is not difficult to directly verify that (\ref{psol1})$\sim$(\ref{psol6}) 
satisfy the relation (\ref{Bianp2}) and the defining equation 
(\ref{gfdef1}) by using the following relations; 
\bea
&&\Delta_i(X,X')=\Delta_i(X',X),\qquad (i=1,2) \\
&&\hat\Delta_i(p,z,z')=\hat\Delta_i(p,z',z),\qquad (i=1,2) \\
&&D_1\left(\partial_z-{{d-1}\over z}\right)
  =\left(\partial_z-{{d-3}\over z}\right)D_2-{2\over z}\sq.
\eea

Having obtained the Green functions in the bulk, we now proceed 
to impose  boundary conditions on them on the brane.  They are 
obtained by matching $\delta(z-L)$ in the defining equation 
(\ref{gfdef1}) and the relation (\ref{Bianp2}); 
\bea
&&2\partial_z\Delta_{\alpha\rho}^T|_{z=L}-c\Delta_{\alpha\rho}^T|_{z=L}=0,
  \label{bch1}\\
&&2\partial_z\Delta_{\alpha\rho}^L|_{z=L}-c\Delta_{\alpha\rho}^L|_{z=L}
  -2\partial_\alpha\Delta_{z\rho}|_{z=L}=0, \label{bch2}\\
&&c\eta^{\beta\nu}\partial_\beta\Delta_{\nu\rho}^L|_{z=L}
  +2M^2\Delta_{z\rho}|_{z=L}=0, \label{bch3}\\
&&c\eta^{\beta\nu}\partial_\beta\Delta_{\nu z}^L|_{z=L}
  +2M^2\Delta_{zz}|_{z=L}=0. \label{bch4}
\eea

It turns out that it is convenient for our 
purpose to set as 
\beq
  c=-{{2(\nu-d/2+1)}\over L}, \label{cvalp}
\eeq
which coincide with (\ref{cval}) when $d=4$.  This value is chosen in 
such a way that the leading pole of $\Delta_{\mu\nu}^T$ is massless 
(see (\ref{fprop1})$\sim$(\ref{fprop4})).  

First we discuss about a solution for $\hat\Delta_1$.  
The boundary condition (\ref{bch1}) is obtained as a consequence of 
the potential on the brane and it differs from the Neumann condition.  
It is a mixed boundary condition of Dirichlet type and of Neumann 
type, which is briefly commented in the previous paper \cite{GN} 
and is employed in \cite{MPi}.  
As we have imposed the boundary condition (\ref{bch1}) on the brane, 
the procedure to obtain the propagator is parallel to \cite{11}.  
The result is given by 
\beq
\hat\Delta_1(p,z,z')={{i\pi L}\over 2}
\left[{{J_{\nu-1}(qL)}\over{H_{\nu-1}^{(1)}(qL)}}
H_\nu^{(1)}(qz)H_\nu^{(1)}(qz')-J_\nu(qz_<)H_\nu^{(1)}(qz_>)\right], 
\eeq
where $z_>$ ($z_<$) denotes the greater (lesser) of $z$ and $z'$.  
A case that is of particular interest here is that where the 
arguments of $\Delta_{\mu\nu}^T(x,z;x,z')$ is on the brane, $z=z'=L$.  
In this case, the propagator is expressed as 
\bea
\Delta_{\alpha\rho}^T(x,L;x',L) 
&=& \int{{d^d p}\over{(2\pi)^d}}e^{ip(x-x')}\left(\eta_{\alpha\rho}
  -{{p_\alpha p_\rho}\over{p^2}}\right){1\over q}
  {{H_\nu^{(1)}(qL)}\over{H_{\nu-1}^{(1)}(qL)}}  \cr
&=& \int{{d^d p}\over{(2\pi)^d}}e^{ip(x-x')}\left(\eta_{\alpha\rho}
  -{{p_\alpha p_\rho}\over{p^2}}\right)
  \left[{{2(\nu-1)}\over{q^2L}}-{1\over q}
  {{H_{\nu-2}^{(1)}(qL)}\over{H_{\nu-1}^{(1)}(qL)}}\right].\label{frop3}
\eea
From this result, some interesting features are observed. Hereafter 
we consider the case of $d=4$.  

As in Ref.~\cite{11}, $\Delta_{\mu\nu}^T(x,L;x',L)$ is split into 
the sum of the 4-dimensional massless propagator and the exchange 
of the Kaluza-Klein states; 
\bea
\Delta_{\alpha\rho}^T(x,L;x',L)
&=& {{2(\nu-1)}\over{L}}\Delta_{\alpha\rho}^{T(0)}(x,x')
  + \Delta_{\alpha\rho}^{T(KK)}(x,x'), \label{fprop1} \\
%\sq\Delta_{\alpha\rho}^{T(0)}(x,x')
%&=& \left(\eta_{\alpha\rho}-{{\partial_\alpha\partial_\rho}\over{\sq}}\right)
%    \delta^4(x-x'), \label{fprop2} \\
\Delta_{\alpha\rho}^{T(0)}(x,x')
&=& \int{{d^4 p}\over{(2\pi)^4}}e^{ip(x-x')}\left(\eta_{\alpha\rho}
  -{{p_\alpha p_\rho}\over{p^2}}\right){1\over{q^2}}, \label{fprop3}\\
\Delta_{\alpha\rho}^{T(KK)}(x,x')
&=& - \int{{d^4 p}\over{(2\pi)^4}}e^{ip(x-x')}\left(\eta_{\alpha\rho}
  -{{p_\alpha p_\rho}\over{p^2}}\right){1\over q}
  {{H_{\nu-2}^{(1)}(qL)}\over{H_{\nu-1}^{(1)}(qL)}}. \label{fprop4}
\eea
The propagator $\Delta^{T(0)}(x,x')$ represents the localized 
massless state in the $4$-dimensional brane.  This localized 
mode is precisely the gauge boson (photon) trapped on the brane.  
This is a consequence of the choice (\ref{cvalp}) as previously noted.  
The leading part of the summation of the KK exchanges gives $1/r^3$ 
potential for $\nu>2$ as in the case of the gravity while it gives 
$1/r^{\nu+1}$ potential for $1<\nu<2$.   Thus we have succeeded to 
trap a photon on the brane as a leading massless mode of vector.  
In order to obtain this result the potential on the brane has played 
an important role.  

On the other hand, boundary conditions (\ref{bch2})$\sim$(\ref{bch4}) 
are satisfied if the following boundary condition for $\hat\Delta_2$ 
is satisfied; 
\beq
\left(\partial_z-{{d/2-1}\over z}-{{2M^2}\over c}\right)
\hat\Delta_2(p,z,z')|_{z=L}=0.\qquad(z'>L) \label{d2bc1}
\eeq
Therefore $\hat\Delta_2$ has a nontrivial solution for the present 
case in contrast to the case when $\Delta_{MN}$ are even functions 
of $y$ for all $M,N=\mu,y$.  In latter case there is no solution 
for $\hat\Delta_2$.  With the value of $c$ given by (\ref{cvalp}), 
the above boundary condition (\ref{d2bc1}) reads as 
\beq
\left(\partial_z-{\nu\over L}\right)\hat\Delta_2(p,z,z')|_{z=L}=0.
\qquad(z'>L) \label{d2bc2}
\eeq
This boundary condition is the same as the one for $\hat\Delta_1$ 
so that 
\beq
\hat\Delta_2(p,z,z')=\hat\Delta_1(p,z,z'). \label{d2sol}
\eeq
Thus $\Delta_2(X,X')$ is split into the sum of 4-dimensional massless 
propagator and the exchange of the Kaluza-Klein states as well as 
$\Delta_1(X,X')$.  However the massless pole of $\Delta_2$ in 
$\Delta_{y\nu}$, $\Delta_{\mu y}$, and $\Delta_{yy}$ cannot be seen 
on the brane because of the factor ${\rm sgn}(y)$ which vanishes 
at $y=0$, just on the brane.
This result is consistent with the consideration 
in the previous section.  

%%%%%%%%%%%%%%%%  Summary  %%%%%%%%%%%%%%%%%%
\section{Summary}
We have examined the massive vector field in the AdS background, 
in which the Randall-Sundrum three-brane is embedded.  
We find the localization of the leading massless mode of the 
transverse part of the massive vector on the brane.  This result 
is assured by both wave-function analysis and propagator analysis.  
The leading massless mode is nothing but a gauge boson (photon) 
as is seen from the effective action (\ref{aeff}).  
In contrast to \cite{Oda}, the mass of the trapped gauge boson 
is strictly vanishing in our analysis.  
For the purpose of obtaining the above conclusion, 
the potential on the brane, which take forms of tachyonic 
mass terms, play an essential role.  Although the origin of the potentials 
on the brane is obscure at present, we would like to point out some 
resemblance of them to the potential on the brane for scalar considered 
in \cite{GW}, where tachyonic mass term is contained in the Higgs-type 
potential.  

On the other hand, it is necessary for getting nontrivial solutions 
for the longitudinal mode ($A_\mu^L$) and the one transverse to the 
brane ($A_y$) to assume opposite transformation properties 
under $y\rightarrow -y$.  If we assumed that $A_\mu^L$ and $A_y$ 
are both even functions of $y$, there would be no nontrivial solutions.  
In the propagator approach, the case of even $A_{\mu}$ and odd $A_y$ is 
examined and the above statement is assured.  
This situation is in contrast to the case of massless vector where odd 
function is not necessary.  

In addition to the potential on the brane, the mass term of the 
bulk vector is necessary to obtain the normalizable wave-function.  
The mechanism proposed here is curious in the sense that the 
broken gauge symmetry in the bulk is restored on the brane.  
It could be considered as a kind of inverse Higgs mechanism 
into the reduced dimensions.  
 
It will be interesting to study the case including the charged 
particles \cite{DRT2,DR}, since our model could provide concrete 
form of z-dependent wave-function and propagator of the gauge fields.  
We will give the discussion related to these in the future.  

%%%%%%%%%%%%%%%%  References %%%%%%%%%%%%%%%%%%


\begin{thebibliography}{99}

\bibitem{RS1} L. Randall and R. Sundrum, Phys. Rev. Lett. {\bf 83} (1999) 3370,
        ({\tt hep-ph/9905221}).
\bibitem{RS2} L. Randall and R. Sundrum, Phys. Rev. Lett. {\bf 83} (1999) 4690,
        ({\tt hep-th/9906064}).
\bibitem{BG} B.~Bajc and G.~Gabadadze, Phys. Lett. {\bf B474} (2000)
        282, ({\tt hep-th/9912232}).
\bibitem{GrN} Y. Grossman and M. Neubert,
        Phys. Lett. {\bf B474} (2000) 361, ({\tt hep-ph/9912408}).
\bibitem{CHN} S. Chang, J. Hisano, H. Nakano, N. Okada and M. Yamaguchi,
        Phys. Rev. {\bf D62} (2000) 084025, 
        ({\tt hep-ph/9912498}).
\bibitem{DS} G. Dvali and M. Shifman,
        Phys. Lett. {\bf B396} (1997) 64
         [Erratum-ibid. {\bf B407} (1997) 452], ({\tt hep-th/9612128}).
\bibitem{DGS} G. Dvali, G. Gavadadze and M. Shifman,
        Phys. Lett. {\bf B497} (2001) 271, ({\tt hep-th/0010071}).
\bibitem{DFKK} P.~Dimopoulos, K.~Farakos, A.~Kehagias and G.~Koutsoumbas, 
        ``Lattice Evidence for Gauge Field Localization on a Brane'', 
        {\tt hep-th/0007079}.  
\bibitem{DLS} M.J. Duff, J.T. Liu, and W.A. Sabra, 
        ``Localization of supergravity on the brane'', 
        {\tt hep-th/0009212}.  
\bibitem{Oda} I.~Oda, ``A New Mechanism for Trapping of Photon'', 
        {\tt hep-th/0103052}.   
\bibitem{GW} W.D.~Goldberger and M.B.~Wise, Phys. Rev. Lett. {\bf 83} (1999)
     4922, ({\tt hep-ph} {\tt /9907447}).
\bibitem{GGH} H. Georgi, A. K. Grant and G.~Hailu, 
        Phys. Lett. {\bf B506} (2001) 207, 
        ({\tt hep-ph} {\tt /0012379}).
\bibitem{GN} R. Ghoroku and A. Nakamura, ``Stability of Randall-Sundrum 
        brane-world and tachyonic scalar'', {\tt hep-th/0103071}.
\bibitem{11} S.B. Giddings, E. Katz and L. Randall, JHEP {\bf 03} (2000) 023,
        ({\tt hep-th/0002091}).
\bibitem{DRT} S.L. Dubovski, V.A. Rubakov and P.G. Tinyakov, Phys. Rev.
         {\bf D62} (2000) 105011, ({\tt hep-th/0006046}).
%%%%%%%%%%
\bibitem{MPi} M. Mintchev and L. Pilo, 
        Nucl. Phys. {\bf B592} (2001) 219, 
        ({\tt hep-th/0007002}).
\bibitem{DRT2} S.L.~Dubovsky, V.A.~Rubakov, P.G.~Tinyakov, JHEP {\bf 08} 
        (2000) 041, ({\tt hep-th} {\tt /0007179}).
\bibitem{DR} S.L.~Dubovsky, V.A.~Rubakov, ``On models of gauge field 
        localization on a brane'', {\tt hep-th/0105243}.

%%%%%%%%%%%%%%%%%%%%%%%%%%%%%%%%%%%%%%%%%%%%%%%%%%%%%%%%%%%%%%%%%%%%%%%%%%
\end{thebibliography}
\end{document}